\documentclass[conference]{IEEEtran}
\IEEEoverridecommandlockouts
\IEEEoverridecommandlockouts
\usepackage{custom_macros} 
\usepackage{cite}
\usepackage{amsmath,amssymb,amsfonts}
\usepackage{algorithmic}
\usepackage{graphicx}
\usepackage{textcomp}
\usepackage{xcolor}
\usepackage{etoolbox}
\usepackage{pgfplots}
\usetikzlibrary{pgfplots.statistics}

\DeclareRobustCommand*{\IEEEauthorrefmark}[1]{%
  \raisebox{0pt}[0pt][0pt]{\textsuperscript{\footnotesize #1}}%
}

\begin{document}

\title{MOAI: A methodology for evaluating the impact of indoor airflow in the transmission of COVID-19}

\author{\IEEEauthorblockN{Axel Oehmichen*\IEEEauthorrefmark{1,2},
Florian Guitton*\IEEEauthorrefmark{1,2},
Cedric Wahl\IEEEauthorrefmark{2}, 
Bertrand Foing\IEEEauthorrefmark{2},
Damian Tziamtzis\IEEEauthorrefmark{2}, and
Yike Guo\IEEEauthorrefmark{1}}
\IEEEauthorblockA{\IEEEauthorrefmark{1} Data Science Institute, Imperial College London, United-Kingdom}
\IEEEauthorblockA{\IEEEauthorrefmark{2} Secretarium Ltd, London, United-Kingdom}
{\{ axelfrancois.oehmichen11, f.guitton, yg \}@imperial.ac.uk}
\thanks{* Those authors contributed equally to this work. This work is supported by Innovate UK (Project Reference 72834)}}

\maketitle

\begin{abstract}
Epidemiology models play a key role in understanding and responding to the COVID-19 pandemic. In order to build those models, scientists need to understand contributing factors and their relative importance. A large strand of literature has identified the importance of airflow to mitigate droplets and far-field aerosol transmission risks. However, the specific factors contributing to higher or lower contamination in various settings have not been clearly defined and quantified. 

As part of the MOAI project (https://moaiapp.com),  we are developing a privacy-preserving test and trace app to enable infection cluster investigators to get in touch with patients without having to know their identity. This approach allows involving users in the fight against the pandemic by contributing additional information in the form of anonymous research questionnaires. We first describe how the questionnaire was designed, and the synthetic data was generated based on a review we carried out on the latest available literature. We then present a model to evaluate the risk exposition of a user for a given setting. We finally propose a temporal addition to the model to evaluate the risk exposure over time for a given user.

\end{abstract}

\begin{IEEEkeywords}
COVID-19, Machine Learning, Exposure Risk, Epidemiology, Privacy, Confidential Computing
\end{IEEEkeywords}

\section{Introduction}\label{sec:intro}
\topic{We explain NPIs are needed while we don't have a vaccine}
Epidemic modelling is vital for understanding the evolving transmission dynamics and responding to the ongoing pandemic. In particular, it is now well established that COVID\nobreakdash-19 transmission commonly occurs in closed spaces. In the absence of a large scale vaccination of the population with an effective vaccine, non-pharmaceutical interventions (NPIs) should be identified and used extensively to reduce potential airborne transmissions.

\topic{We explain that various CDCs (EU, UK, US) do mandate NPIs to be put in place, but conclusive proofs are needed to guide those policies}
A variety of NPIs have been adopted worldwide based on Centres for Disease Prevention and Control (CDCs) recommendations. In common with many other countries, the UK implemented a mass ``stay-at-home'' order; non-essential stores have been closed, mandatory face masks in stores must be worn, one way systems in stores have been put in place, etc. All those NPIs are educated guesses based on prior viral outbreaks and scientific knowledge on known transmissions vectors for airborne viral diseases. That lack of knowledge is illustrated by the sometimes divergent approaches countries have taken to mitigate the propagation. For example, children's role in school settings resulted in many shifts in some countries (e.g. UK,  Ireland, Germany or Denmark) of opening and closing over time in various countries based on evolving scientific consensus. Some NPIs have proven counterproductive as well, even though it might appear counter-intuitive, such as one way systems~\cite{Ying2020ModellingModel}. Nevertheless, in the absence of a precise understanding of the transmission mechanisms of COVID\nobreakdash-19, the NPIs put in place are fairly broad and coarse, which puts more pressure on our economies, mental health and personal liberties. In order to alleviate that pressure, a more precise understanding of closed spaces exposure risks is necessary. \\
\topic{We explain that fluid mechanics modelling is necessary but only part of the equation toward identifying the contributing factors}
The modelling of airborne transmission in indoor spaces fundamentally relies on the fluid mechanics modelling of the transport of viral aerosols. In order to achieve accurate modelling, understanding contributing factors is paramount. In general, models studying airborne transmission risk fall into one of two types: those based on the assumption of well-mixed room (WMR) and models of computational fluid dynamics (CFD). The WMR assumption is that virus-carrying aerosols are spread uniformly across the room instantaneously\cite{Riley1978AirborneSchool, Wells1955AirborneInfections}, so that everyone in the room, regardless of their position, is equally likely to be infected. By ignoring the complex effects of airflow on the airborne particles, this assumption greatly simplifies the problem, so the models can be incorporated into spreadsheets and are swift to run. That simplicity and lightweightness allowed us to gain some insights on COVID\nobreakdash-19 propagation~\cite{Dai2020AssociationInvestigation, Buonanno2020EstimationAssessment, Sun2020TheTransmission}. Various CFD simulations have also investigated the transport of viral COVID\nobreakdash-19 aerosols \cite{Vuorinen2020ModellingIndoors, Shafaghi2020OnTransmission, Feng2020InfluenceStudy,Li2020EvidenceRestaurant,Birnir2020TheSpaces, Shao2021RiskSettings}. CFD  simulations are powerful tools to study airborne transmission as they allow for fine-grain details such as the room size, geometry, the complex turbulent airflow (e.g. open windows or HVACs) and the size distribution of the aerosols, such as in \cite{Vuorinen2020ModellingIndoors}.

Nevertheless, those models are only as accurate as the quality of the contributing factors considered. As part of the MOAI project, we aim to identify those factors systematically and propose a simpler approach to model the exposure risk for all settings by only considering contributing risk factors coupled with proven exposures from the NHS Test and Trace program and not the airflow itself. To identify those factors, we developed a user-based questionnaire integrated into the MOAI app that would allow users and venues to contribute a more accurate description of the airflow in the setting. We then built an ensemble model which evaluates the likelihood of exposure of any new QR code scanning into a venue by a user.

\topic{The model is not the only challenge. Privacy and user experience must be at the heart of the platform for a successful adoption}
Lastly, citizens’ concerns about data privacy and data security breaches may reduce the adoption of COVID\nobreakdash-19 contact tracing mobile phone applications, making them less effective~\cite{Horvath2020CitizensApps}. Studies~\cite{Horvath2020CitizensApps, OCallaghan2020AIreland, Zhang2020AmericansPandemic} conducted in different countries have highlighted the importance of centralised National Health Service systems and a mixture of digital and human contact tracing for a successful adoption and trust by the public. 

The contributions of this paper are the following:
\begin{itemize}
\item We present MOAI's modular and scalable architecture, designed with a combination of security and privacy features that allows contact tracing and risk of exposure evaluation while preserving the participating users' privacy (\sref{sec:moaiapp}). 
\item We detail the literature review conducted, which lead to identifying potential contributing factors, the design of the questionnaire, and the generation of the synthetic exposure data (\sref{sec:data}).
\item We propose an exposure risk model used by the platform to provide feedback to the users on their current exposure risk (\sref{sec:exposuremodel}).
\end{itemize}



\section{MOAI: Federated Data Collection and Collaborative Research}\label{sec:moaiapp}

\topic{A brief overview of the architecture and its components.}
In this section, we provide an overview of the architecture of the MOAI platform. The platform is divided in three main parts: the mobile application, the web portal and the Secretarium back end. The exposure data from the questionnaires is collected centrally from the app to the Secretarium back end in an encrypted fashion for storage and compute the risk of exposure in enclaves using the proposed model. The Secretarium back end provides secure public APIs for researchers to interact with the platform and secure web services for health authorities' teams to discuss with patients in the eventuality of a confirmed contact (e.g. Test and Trace teams in the UK).

\topic{This section we dissect the rationale behind the design approach.}
Building software for NPIs initiatives can be a delicate political, privacy and ethical exercise. Recommendations provided by health protection agencies have not only been criticised by academic experts~\cite{LI2020,Guinchard2020}, there are multiple issues of engagement sprawling from distrust from the general public \cite{Guinchard2020}. To make the matter worse, most recommendations are region-centric in their legal framework and implementation, thus preventing more efficient cross-nation collaborative efforts. A notable exception can be made of PEPP-PT \cite{Cooper2020} but it was met with a large amount of controversy \cite{SEBASTIANKLOCKNER2020} leading to more academic driven initiatives such as DP-3T \cite{Avitabile2020}. The lack on consensus led to the creation of many different separate applications \cite{Garousi2020} and left governments only with stringent travel restriction strategies to stop the spread across borders. This appears to be among the few measures that countries all  commonly implemented; with the drawback of hindering people's freedom and hugely impacting economies and mental health issues in large swathes of the population.

\subsection{The Secretarium platform}
\topic{We present the architecture of MOAI at high level}
The Secretarium platform is the fruit of Blockchain, Trusted Execution Environments (TEEs) and cryptography research and developments. We leverage secure multi-party computing techniques~\cite{Yao1986HowSecrets} and TEEs to achieve strong security and privacy guarantees when data is at rest, in transit and in use with multiple untrusted (potentially adversarial) parties. Secretarium's platform enables 'Code-Is-Law' implementation and is intended to satisfy some important properties: 

\textbf{Secure.} The platform and data must be secure against penetration attacks aiming to gain unauthorised access to data. The Secretarium Communication Protocol (SCP) guarantees that private and confidential data stays under the control of their originator at all times. We follow the principle of end-to-end encryption throughout the data's life cycle; none of Secretarium engineers or other node runners get access to the data in clear-text.

\textbf{Integrity-preserving.} The platform's results must be tamper-proof, including the party performing the analysis (e.g. Researchers) and must provably yield results as specified. This mechanism is similar to the tamper-proof integrity mechanisms available on a Blockchain. The MOAI application within the Secretarium platform is hosted on an internet-like public network. We have developed a Byzantine Fault Tolerant version of the RAFT algorithm~\cite{Ongaro2014InAlgorithm} to allow the application to run in a distributed fashion without a single point of failure.

\textbf{Privacy-preserving.} Queries results should consist of aggregated data or simple yes/no answers and never disclose individuals private information stored in the ledger. In the context of this project, the data available to the researchers contains no personally identifiable field and the timestamp is coarsened to fixed four hours windows (e.g. 0-4, 4-8, ..., 20-24).

\textbf{Flexible \& Scalable.} The platform must support a large array of purposes. Data coupled with each application must be segregated to maximise performance. The platform must be multi-layered, modular, and scalable. The modularity of the architecture and flexibility in the layers allows for horizontal scaling of each layer independently and enables decoupled upgrade, addition and removal of services with no downtime.

\textbf{Simplicity.} Secretarium's protocol is designed to facilitate integration with existing systems, especially to allow recent commonly used internet browsers to serve natively as clients.

\begin{figure}[!ht]
\centering 
\includegraphics[width=0.49\textwidth]{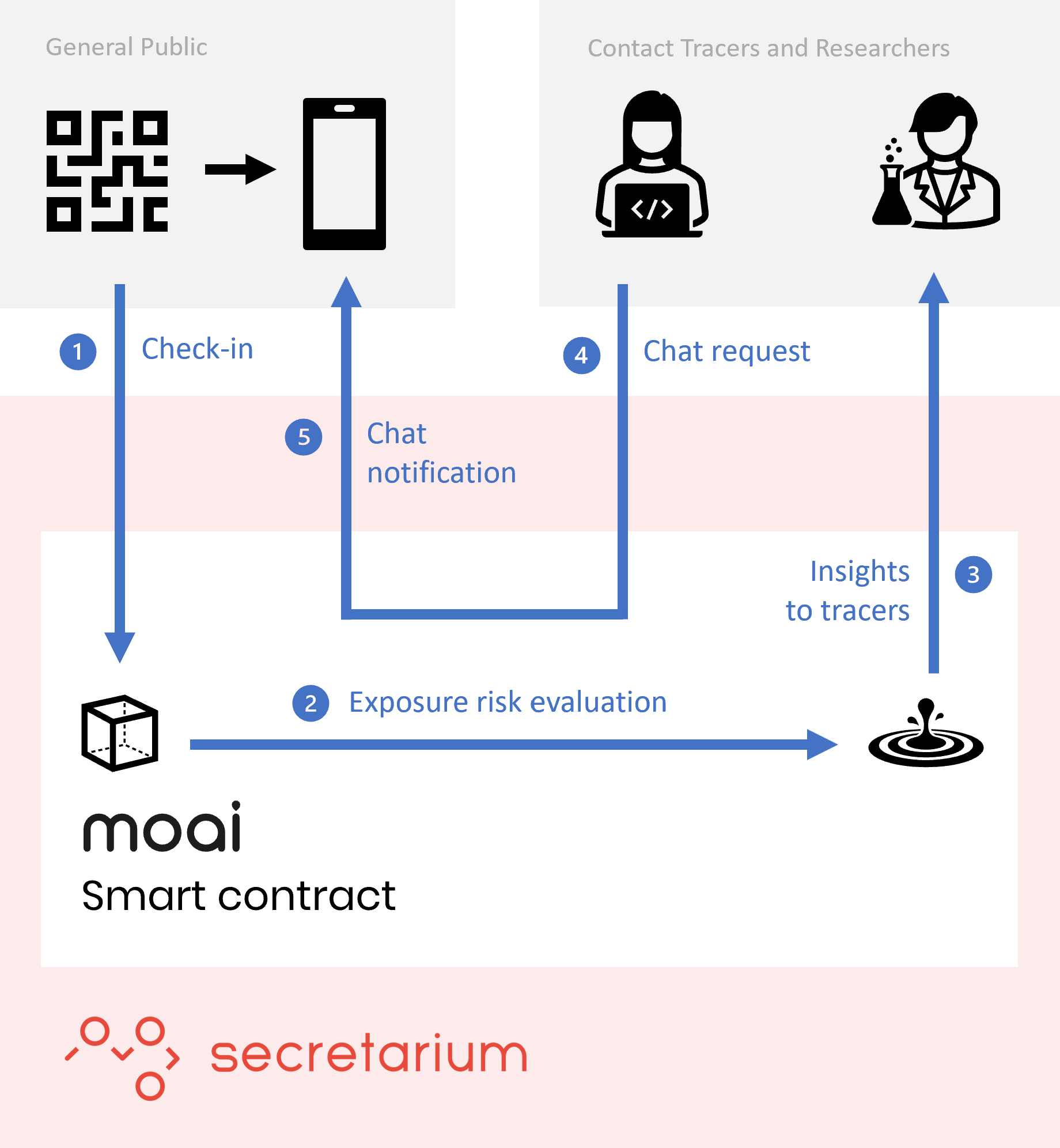}
\caption{\label{fig:MOAI}A high-level schematic representation of the data flow of the MOAI platform. The pale red Secretarium container represents a single enclave, many are deployed to provide MOAI with sufficient fault tolerance and performance.}
\end{figure}

\subsection{Security, privacy and compliance considerations}
MOAI's first and foremost goal is to design a solution that does not infringe on participants' privacy, be it of the businesses and venues or the general public. Indeed, the large adoption in the UK of the NHS COVID\nobreakdash-19 app (more than 5 million installations for Android alone as of March 2021) and the implication of large numbers of volunteers for the Test and Trace teams creates a large surface of attack for malicious insiders aiming at gaining some information about users. For this reason, MOAI is not collecting any personally identifiable data for it to provide a viable functional level of operation. Data collected in the context of MOAI research questionnaires, while optional, are treated and stored in the most secure fashions relying on confidential computing leveraging Intel SGX enclaves. Best practices in auditing and data provenance and usage tracking can also be ensured by Secretarium technology's guarantees on top of the Intel SGX platform. By leveraging those guarantees, MOAI is compliant with the requirements posed by regulations such as GDPR and HIPPA by design.

\subsection{Hardware Compatibility}
\topic{Discuss the issue of Exposure Notification APIs and other Bluetooth based API.}
Our approach took into consideration limitations pertaining to participants devices. Indeed, many implementations of such software NPIs relied on either utilising Bluetooth beaconing techniques or what came to be known as the Exposure Notification API (a collaboration between Apple and Google to provided a cross-platform solution to the contract tracing requirements of multiple government bodies \cite{Leith2020}). In addition to technical limitations (e.g. no guarantee that two people in relative proximity are in the same room), both these approaches have hardware and platform constraints that limited their effectiveness. Eventually, projects like the NHS's COVID\nobreakdash-19 mobile application pivoted to using QRCode-based contract tracing, increasing the eligible hardware considerably. This change, compounded with later terms and conditions of use changes for the Exposure Notification API brought by Apple and Google \cite{FutureofPrivacyForum2020}; prompted us to envisage a similar solution based on QRCodes.

\subsection{NHS compatibility}
While providing a robust privacy-preserving solution for contact tracing, another central goal of MOAI is to increase cross-agency collaboration. Our first step was to design a solution that would extract information from existing ``NHS-COVID\nobreakdash-19'' QRCodes so that a large amount of existing QRCodes can be leveraged. These QRCodes contain a colon-separated set of clear text and base64 encoded RFC7515 JSON Web Signature (JWS) object \cite{Bradley2015}. MOAI extracts the ``id'' and ``vt'' members of the payload particle of the JWS data. ``id'' referring to the unique identifier provided to the venue and ``vt'' denoting its type. The venue type is then mapped to the types of venue available in MOAI, naively extending the NHS list (See Appendix \ref{annex:NHSLocations}). The ID is ingested and treated as immutable and unique. MOAI does not validate the signature of the token from the "NHS COVID\nobreakdash-19" QRCodes as it does not have the necessary private cryptography components and the veracity of the ID is not relevant in this context.

\subsection{Scalability}
\topic{Describe the strategy implemented for the scalabilty of the platform }
Relying on a ledger-based technology, even when powered by data replication and load balancing, generates a significant performance bottleneck when scaling to support hundreds of thousands of simultaneous queries (new QR code scans by the users, test and trace teams, chats, etc.). MOAI overcomes that limitation by pooling data capture in ledger silos and allowing cross-silo querying via a federated query layer.
Nevertheless, every silo follows the best practice of data replication across multiple enclaves and leverages particular implementation of consensus mechanism using a variant of the
Raft algorithm capable of achieving Byzantine Fault Tolerance. This allows a total ordering of transactions across a cluster, necessary to maintain transactional integrity in the eventuality of the failure of any component of the system.  This approach allows as well to overcome any imperfect information on whether a component has failed \cite{Lamport1982}.

\subsection{Secure transmission and storage}
\topic{Describe the cryptographic pipeline used for collecting and querying information}
The transfer of any data to and from MOAI's TEEs is done via the internet using the Secretarium Connection Protocol (SCP). The SCP is a variant of the Transport Layer Security (TLS) 1.2 protocol~\cite{Dierks2008The1.2}.
As TLS, SCP guarantees a secure channel permitting authentication, privacy and integrity/reliability of the communication. In addition, SCP provides three significant gains:
\begin{itemize}
	\item It guarantees to end-users what application will be the recipient of the data on the remote machine.
	\item It guarantees privacy and integrity against an adversary with direct access to the remote machine, by providing cryptographic proofs that the application runs in a trusted execution environment.
	\item It provides protection against denial of service attacks through.
\end{itemize}
To achieve a degree of protection against bruteforce denial of service attacks (DoS) SCP relies on adaptive difficulty proof-of-work challenges. After the client advertises its ephemeral public key for the connection the server will decide of a difficulty level for a proof-of-work calculation the client has to compute. Multiple things can be included in the calculation of the difficulty level such as the number of connection the client attempts in a short period of time, the number of time a particular route is accessed or how much payload is being send over the connection.

\begin{figure}[!ht]
\centering 
\includegraphics[width=0.49\textwidth]{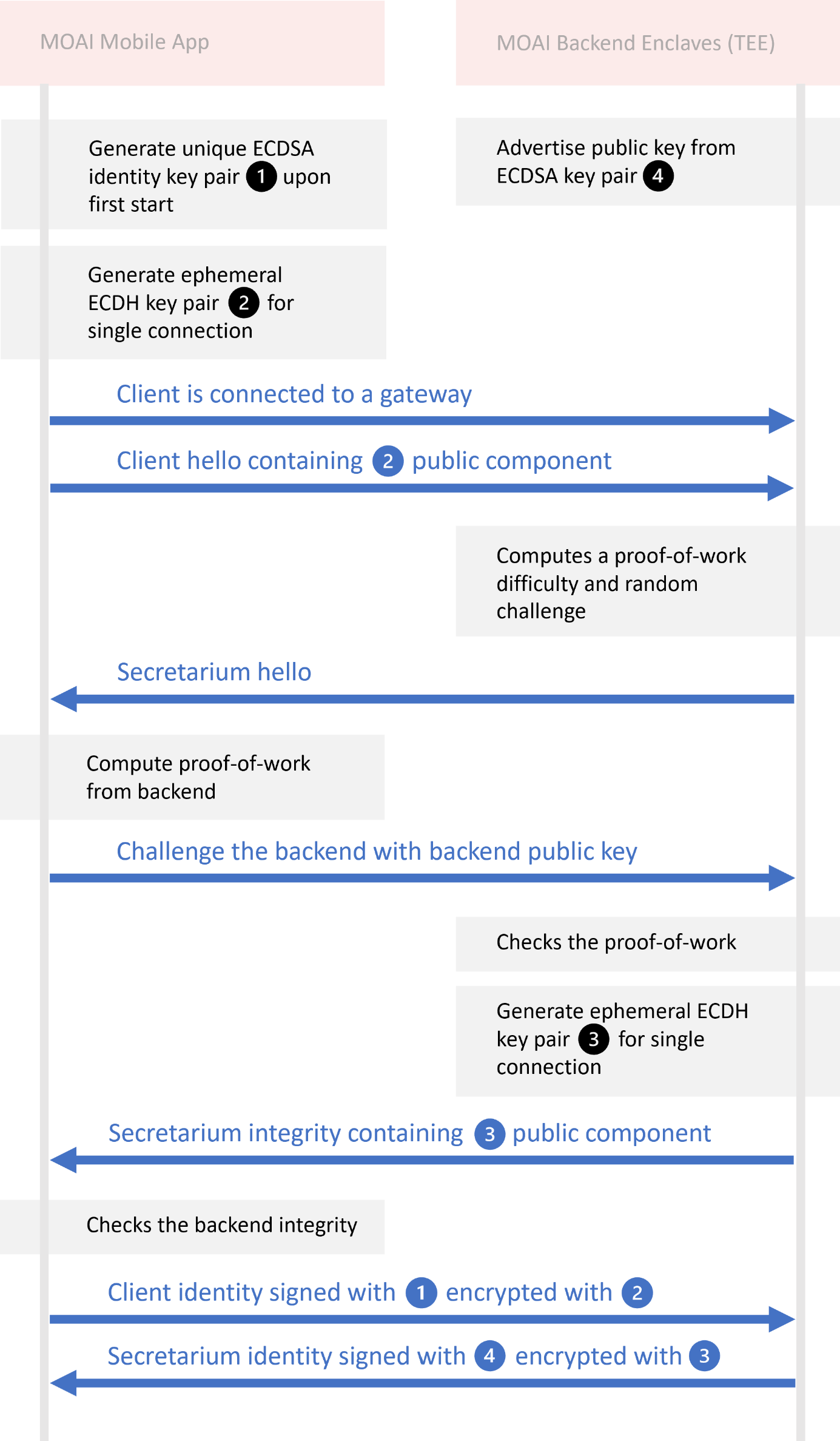}
\caption{\label{fig:SCP}A high-level schematic representation of the SCP protocol communication between the MOAI App and the MOAI back end.}
\end{figure}

\subsection{Federated Querying}
\topic{Present the tools developed to permit distributed querying of the data by representatives and healthcare volunteers}
\topic{Present as well the privacy layer with the fact that the data is processed in enclaves and no identifiable data is ever extracted for building the exposure model}
MOAI proposes a web portal meant to be used by healthcare workers and health authorities representatives to help identify and communicate with participants potentially at risk. MOAI does not offer direct access to its users' related data. Instead, it requires the investigator to enter the venue's unique identifier and a date and a time for the search. A set of identifier results from the search, and the investigator can directly start a written communication channel with the user on the other end. In ensuring compatibility with the ``NHS COVID\nobreakdash-19'' QRCodes, we permit the search to be done via the NHS unique venue ID only as MOAI does not extract the location information encoded into QRCodes.

\subsection{Technology stack}
The back end of MOAI relies on a C++ Smart Contract which runs within a Secretarium enclave on top of a Windows Server. The mobile application and portal are written in TypeScript using the ReactJS library. The mobile app is packaged and made cross-platform Android and iOS via the use of Expo.io. The front end portal is a standard HTTP server on a linux machine.

\subsection{Exposure level}
\begin{figure}[!ht]
\centering 
\includegraphics[width=0.49\textwidth]{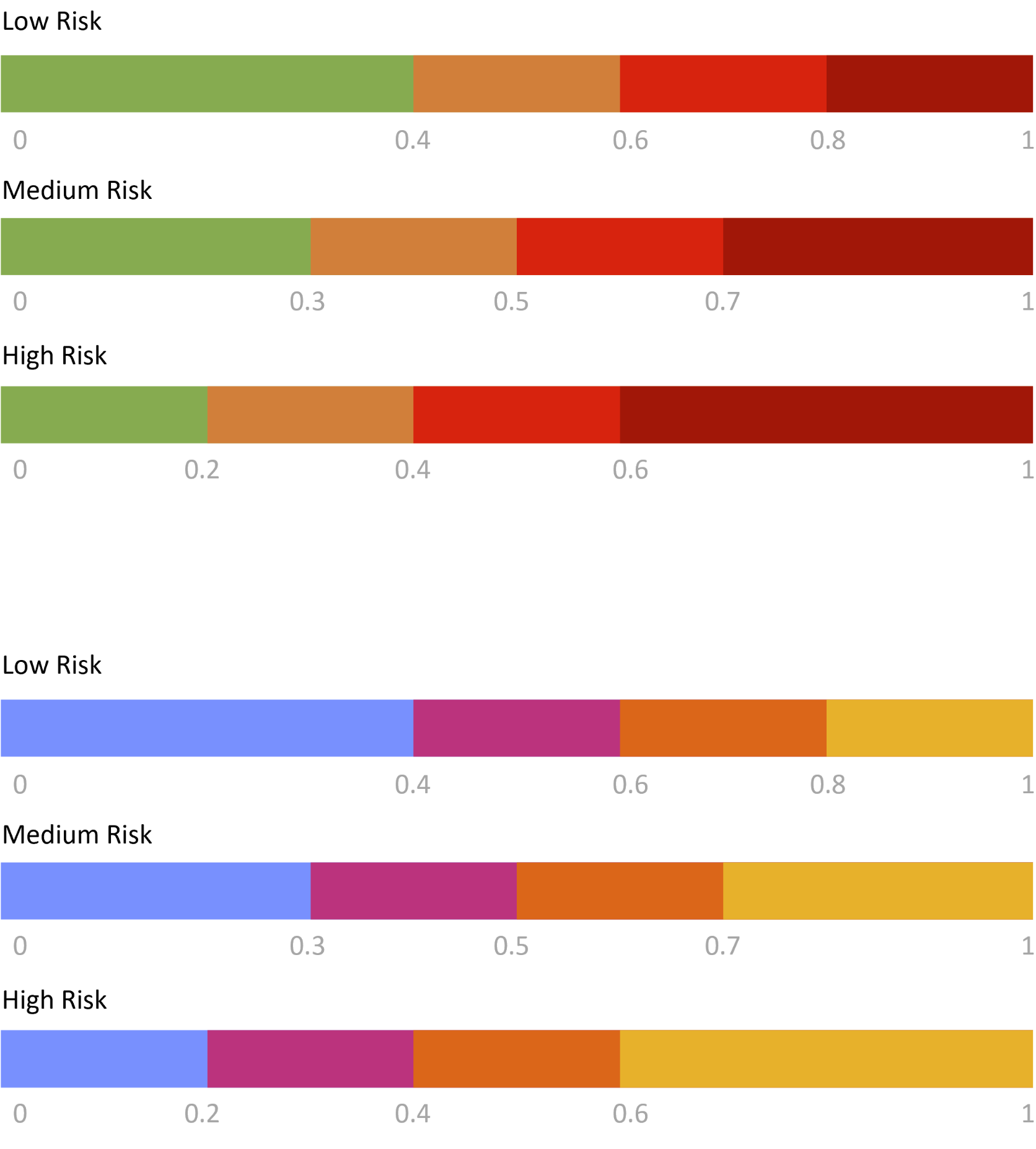}
\caption{\label{fig:risk_levels} Risk levels with their associated threshold for the three categories of users. At any given point in time, the users fall into on of the four categories with their score (e.g. their risk of exposure) between 0 and 1 being computed using our proposed model. Also displayed the adapted colour-scale for visually impaired users.}
\end{figure}

\topic{we explain that we do not display the raw exposure risk for four levels of exposure that are computed using the model and modulated using the inputted risk factors someone with higher risks as defined by the NHS while have lower thresholds for the risk factor to reflect better to the users}
The raw exposure risk would be difficult to understand for most users as it doesn't translate in a simple to understand notion for them. Thus, we are translating this risk into a simple four colour coded categories: low, medium, high and very high. In addition to the improved user experience in the app, the thresholds can be modulated per user based on their profile risk using the users inputted risk factors. Those risk factors fall into the three categories defined by the NHS and detailed in Annex \ref{app:Riskfactors}. This approach allows us to send a clear real-time feedback to the user with respect to their exposure behaviour and hopefully positively influence their attitude toward exposure risk. Figure \ref{fig:risk_levels} details the variable threshold for the different categories of users. Because the app is designed to be used by a broad population, we propose an option to replace the red/green palette with a blue/orange one to be more colour-blind-friendly. This option might seem anecdotal but there are approximately 3 million colour blind people in the UK alone (about 4.5\% of the entire population).

\section{Questionnaire design and Synthetic Exposure Data Generation}\label{sec:data}

\subsection{Identified exposure risks from the literature}
\topic{Main transmission of COVID\nobreakdash-19 and use usefulness of PPE}
Currently available evidence indicates that COVID\nobreakdash-19 may be transmitted from person to person through several different routes. Infection is understood to be mainly transmitted via large respiratory droplets containing the COVID\nobreakdash-19 virus. In the review published by La Rosa et al.~\cite{LaRosa2020CoronavirusReview}, COVID\nobreakdash-19 primarily transmits on a person-to-person contact mode through respiratory droplets generated by expectoration (e.g. breathing or sneezing), as well as contact (direct contact with an infected subject or indirect contact from contaminated fomites). Determining particular places linked to clusters of cases could reveal settings and factors responsible for amplifying the heterogeneity in transmission reported: potentially 80\% of transmission is being caused by only 10\% of infected individuals~\cite{Endo2020EstimatingChina}.

Multiple outbreaks and clusters of COVID\nobreakdash-19 have been observed in a variety of indoor settings have been reported since the start of the pandemic in the European Union, the European Economic Area and the United Kingdom. Those combined countries reported 1377 clusters~\cite{Ecdc2020COVID-19UK} of COVID\nobreakdash-19 in occupational settings which occurred between March and early July 2020. The main potential factors contributing to clusters and outbreaks in occupational settings identified so far are:
\begin{itemize}
\item \textbf{Working in confined indoor spaces}: Studies have shown that in Europe $>$80\% of working time is spent indoors. Participating in meetings and sharing the same office space has been reported in the literature as a risk factor for contracting COVID\nobreakdash-19~\cite{Park2020CoronavirusKorea, Rothe2020TransmissionGermany}. Sharing facilities (e.g. canteen and dressing rooms ), transports and accommodations are likely contributing to transmission~\cite{CenterforDiseaseControlandPrevention2020MeatEmployers}.
\item \textbf{Close/direct contact with COVID\nobreakdash-19 cases}: Many essential workers are client-facing or near other coworkers and thus at greater risk of occupational exposure than the rest of the population. Healthcare workers are known to be at substantial risk of occupational exposure to biological agents. Based on UK Biobank data, Mutambudzi et al.~\cite{Mutambudzi2020OccupationParticipants}
has shown that the risk of healthcare workers testing positive for COVID\nobreakdash-19 was over seven times higher than for non-essential workers, and those in social care had a three times higher risk.
\item \textbf{Insufficient or incorrect use of protective personal equipment (PPE)}: Many clusters and outbreaks were linked to either insufficient access to PPE~\cite{Zhang2020ProtectingSuggestions} or inadequate infection control and hygiene standards~\cite{Baker2020Nonrelocatable2018}. Studies~\cite{Suzuki2021EffectivenessWorkers, Wang2020PotentialWorkers, Liu2020UseStudy} have demonstrated that the use of PPE and physical distancing are successful at preventing infection even among the most at-risk group (e.g. healthcare workers).
\item \textbf{``Presenteeism''}: Impossibility to work remotely coupled with financial constraints leading to continued commuting and working, even when the employee is exhibiting COVID\nobreakdash-19 symptoms~\cite{Baker2020Nonrelocatable2018}.
\item \textbf{Transmission from asymptomatic people}: The absence of symptoms in infected people can arise from two different infection states: presymptomatic individuals and individuals who never experience symptoms. It has been demonstrated that asymptomatic persons could transmit COVID\nobreakdash-19 to others~\cite{Oran2020PrevalenceReview} and that transmission from asymptomatic individuals was estimated to account for more than half of all transmissions~\cite{Johansson2021SARS-CoV-2Symptoms}.
\end{itemize}

One last element that was left out of this list is the potential influence of the airflow topology, in a given setting, on transmission of COVID\nobreakdash-19. For the same type of setting and identical number of people, the humidity may influence viral transmission by affecting how droplets/aerosols move and their rate of decay, while the use of Heating, Ventilation and Air-Conditioning systems (HVACs) may have a complementary role in decreasing potential airborne transmission of COVID\nobreakdash-19.

\topic{Quality of the air flow with respect to humidity}
Earlier research indicates that humidity can influence individuals' susceptibility to infection by affecting how droplets and aerosols move and their rate of decay. Droplets are usually categorised as larger entities ($>$5 $\mu$m) that rapidly drop to the ground due to gravity, while aerosols are smaller particles ($\leq$ 5 $\mu$m) that rapidly evaporate in the air. Aerosols leave behind droplet nuclei that are small enough and light enough to remain suspended in the air for hours. According to Zhao et al.'s model~\cite{Zhao2020COVID-19:Droplets} droplets can travel three times farther in low-temperature and high-humidity environment, whereas the number of aerosol particles increases in warm, dry environments. This research also highlighted the importance of proper ventilation, as droplets and aerosol spread significantly farther in air streams. In addition to the influence on the spread of the droplets, a lower ambient humidity impacts our respiratory immunity significantly by diminishing cilia’s capability to expel viral  aerosols~\cite{Kudo2019LowInfection}. The human ear, nose and throat areas are more effective as virus fighter at higher ambient humidity values rather than when room air is very dry~\cite{Ahlawat2020AnEnvironments}. 

\topic{Influence of the use of HVAC in closed spaces}
The importance of controlling quality (including humidity and flow) can be achieved through different means, among them HVACs. Some guidance around this have been published by the various CDCs (UK, EU and US)~\cite{ScientificAdvisoryGroupforEmergencies2020RoleSAGE-EMG, NationalCenterforImmunizationandRespiratoryDiseasesNCIRD2020VentilationCDC, EuropeanCentreforDiseasePreventionandControl2020HeatingUpdate} to guide public policy. Those documents also highlighted the known unknowns with respect to exposure risks in confined indoor spaces. Among those unknowns, these documents have highlighted the importance to improve ventilation in multi-occupant spaces especially if temperature and humidity are low ( risks of higher far-field aerosols as explained before) and include activities that generate high levels of aerosol (e.g. singing or physical exercise). 

\topic{We transition to the questionnaire}
While assessing precisely the quality of airflow in a given setting requires a lot of time and expert knowledge, it is possible to assess it by proxy by leveraging Large scale users-contributed observational data coupled with exposure investigations carried out by the test and trace teams. That approach could enable the collection of data that would otherwise be extremely lengthy and costly to collect.

\subsection{Questionnaire}
\topic{We detail the questionnaire}
A short and targeted questionnaire, embedded in an app, is a straightforward means to collect data such as nature of the building, density of people, ventilation type, length of exposure and activity. The answers to the questionnaire could be contributed by both the app users and venue owners anonymously.

\topic{Anonymity of the questionnaire and collection of the data}
The questionnaire does collect granular timestamps (day and time range for the hour) from the scanning into a venue but at the granular level to mitigate any geolocation re-identification. Furthermore, each venue has a unique random ID in place of location data to preserve privacy. 

To avoid any free text input, the questionnaire only contains multiple choices questions which lowers the time needed by the users to fill up the questionnaire, avoids invalid inputs and enables more accurate modelling. This more straightforward approach is inspired by recommendation systems and Netflix's change from a star rating to like/dislike approach.

The final exposure data contains 22 fields (see Table \ref{table:exposure_structure_table}) that forms the basis for the training of the machine learning algorithm for the platform.

\begin{table}[ht!]
\centering
\begin{tabular}{l|p{4.5cm}}
\hline
\textbf{Field} & \textbf{Definition}\\\hline
TIMESTAMP &  Timestamp of the QRCode scan\\
UserID  &  Unique random ID associated with a user \\
Location\_Type &  NHS defined location types \\
Location\_Inside\_or\_Outside &  Clarify if location is indoor or outdoor\\
Number\_of\_People\_Present &  Estimated number of people present in the location\\
Time\_Spent\_on\_Location &  Time spent at the location\\
Wearing\_Masks &  If people were wearing masks\\
Staff\_Properly\_Wearing\_PPE &  Staff properly wearing their PPE\\
People\_Properly\_Wearing\_PPE &  People properly wearing their PPE\\
Social\_Distancing &  Was the social distancing rules followed\\
Additional\_Measures\_in\_Place &  Additional protective measures implemented at the place\\
Number\_of\_People\_in\_the\_Party &  Number of people in the party\\
All\_Members\_of\_Household &  Members were part of the household\\
All\_Members\_of\_Support\_Bubble &  Some members were not part of the household but the bubble\\
Quality\_of\_the\_Airflow &  Estimation of the quality of the airflow\\
Temperature\_in\_Venue &  How was the relative temperature in the venue\\
Humidity\_in\_Venue &  How was the relative humidity in the venue\\
Clean\_after\_Every\_Usage &  Was the location cleaned after usage\\
Any\_Contact\_Between\_Members &  Any close contact with members of the party (handshake, etc.)\\
Physical\_Activity &  Was the activity a physical or involved extensive expectoration\\
Exposure\_Led\_to\_Contamination &  Outcome of the exposure (validated by a test)\\
Risk\_of\_Contamination & Risk of exposure\\
\hline
\end{tabular}
\vspace{0.1cm}
\caption{Structure of an exposition record. All the fields are categorical variables with the exception of the TIMESTAMP which is a date and UserID which is a random UUID generated for a given user.}
\label{table:exposure_structure_table}
\vspace{-0.5cm}
\end{table}

The full questionnaire is available in Appendix \ref{app:UsersQuestionaire}.

\subsection{Synthetic data}
\topic{We explain how we generated the synthetic data from the questionnaire}
In the context of this research, the design of the synthetic data aims at reflecting the available contamination data from the literature. In the absence of ground truth for the modelling, the identified factors cannot be easily quantified independently for the synthetic data. Therefore, we propose approximating it by modelling it in a similar fashion to a Hidden Markov Model (HMM) to estimate the exposure risk outcome. In this context, the observed outcomes (positive outcome for an exposure) with multiple hidden states (e.g. type of setting, airflow, etc.). The search strategy to find publicly available information on the contamination in different settings is inspired by Leclerc et al.~\cite{Leclerc2020WhatClusters}. 

The exposure risk baseline for all settings is set at 10\%. That baseline is inferred from the statistics published by DHSC, NHS Test and Trace statistics~\cite{Foundation2021NHSTracker}. In addition to that baseline, a modulation based on the identified risks from the literature and a random noise value drawn from a Laplace distribution $Laplace(\mu=0,b=0.5)$ is added.

Each question from the questionnaire is translated into a categorical variable as the inputs as clearly defined and restricted. In the future, in the eventuality of a collection of real data, this would also allow us to display various statistics to the researchers where each data point represents a scan by a user coupled with a COVID\nobreakdash-19 test outcome (positive or negative contamination).

\section{Exposure risk evaluation}\label{sec:exposuremodel}
In the absence of ground truth and actual data, the following evaluations aim at providing an indication and groundwork for future work. In the eventuality where actual data is collected, we will re-evaluate those proposed models again and confirm the validity of the proposed methodology.

\subsection{Experimental design}

To avoid potential unwanted side effects from the random generation of the synthetic data, we generated twice five different data sets containing 150k, 250k, 500, 750k and 1M records, respectively. Then, we split the generated data set into a training set containing 70\% of the data and a test set containing the remaining 30\%. We tuned the hyperparameters using the training data sets and applied cross-validation with k-folds within this sub data set, where k was set to 10. Specifically, in each fold, we used recordings from $N_r$ -- $(N_r/k)$ to train the model, and from the remaining $N_r/k$ subjects to test the trained model, where $N_r$ is the number of records in the data set. This process was repeated k-times so that all of the recordings were tested. The hyperparameter tuning and calibration of the models was done using a random grid search that optimised primarily the accuracy of the model and secondary the F1--score.
The test set was used to compute the final performances of the proposed models and validating the choice. All computed the performance metrics will be discussed in \ref{sec:model_selection}.

In order to check the predictive capability of our selected features, we have selected and tested different classification algorithms, including Logistic Regression, Naive Bayes, Ridge Regression~\cite{Tikhonov1963SolutionMethod}, Ada Boost~\cite{Freund1997ABoosting}, Linear Discriminant Analysis, Random Forests~\cite{Ho1995Random1}, Light Gradient Boosting Machine~\cite{Ke2017LightGBM:Tree}, Gradient Boosting Classifier~\cite{Friedman2001GreedyMachine} and Extreme Gradient Boosting~\cite{Chen2016XGBoost:System}.
\subsection{Implementation}

During the exploration and discovery phase, we extensively relied on PyCaret~\cite{MoezAli2020PyCaret:Python} to evaluate the best models. This library allows us to speed up the code writing, training, tuning and validation of the different models by delegating the tasks to the CPU or GPU (whenever available) seamlessly and providing a single programmatic interface to all the models. The training time for each training fold ranged from a few hundred milliseconds to approximately five minutes on the node for the 1M records data set. The tuning of the top 5 best performing models during training would take less than 2 hours for the 1M records data set.

\begin{table*}[!ht]
\renewcommand{\arraystretch}{1.3}
\centering
\begin{tabular}{|c|c|c|c|c|c|c|c|}
\hline
Model &	Accuracy & AUC & Recall & Precision & F1 & Kappa & MCC  \\ \hline
Light Gradient Boosting Machine & 0.697 &	0.755 &	0.749 &	0.651  & 0.697  & 0.354  & 0.358  \\
Logistic Regression &	0.697  & 0.755  & 0.743  & 0.653  & 0.695  & 0.354  & 0.357  \\
Linear Discriminant Analysis &	0.696  & 0.755  & 0.706  & 0.664  & 0.684  & 0.354  & 0.355  \\
Ridge Classifier &	0.691  & 0  & 0.706  & 0.664  & 0.684  & 0.354  & 0.355  \\
Ada Boost Classifier &	0.676  & 0.756  & 0.783  & 0.642  & 0.705  & 0.354  & 0.363  \\
Gradient Boosting Classifier &	0.676  & 0.754  & 0.820  & 0.634  & 0.715  & 0.354  & 0.370  \\
Extreme Gradient Boosting & 0.675  & 0.754  & 0.705  & 0.662  & 0.683  & 0.351  & 0.352  \\
Random Forest Classifier &  0.648  & 0.705  & 0.631  & 0.649  & 0.640  & 0.297  & 0.297\\
Naive Bayes &	0.618 & 0.721 &	0.318 &	0.782 &	0.452 &	0.232 &	0.288 \\ \hline
\end{tabular}
\caption{\label{tab:performances}Comparison of the performances between the different models evaluated}
\end{table*}

\subsection{The proposed exposure model}

Our proposed model to measure the risk exposure for a given individual at any given time t is as follow:

\begin{equation} P(NotExposed)(t) = 1 - \prod_{i=1}^{n}{(1 - P_{n}(Exposed) e^{-\lambda \delta(t, t_{n})} )} \end{equation}

where P is the probability of exposure at the current time time t, n is the number of exposures (e.g. number of scans done by the user at a setting), $P_{n}$ is the probability of exposure at the given setting and $e^{-\lambda \delta(t, t_{n})}$ the weight decay with $t$ the current time, $t_{n}$ the time when the exposure occurred, $\lambda$ the decay constant and the function $\delta$ defined as:

\[
\delta(x,y) = 
\left\{
     \begin{array}{@{}l@{\thinspace}l}
       \text{0}  & \text{ if (x - y)} \le \text{2 days}\\
       \text{(x - y - 2880)} & \text{ if (x - y)} > \text{2 days} \\
     \end{array}
   \right.
\]

The function $\delta$ is a function of time in minutes where 2880 represents the number of minutes in two days. In the context of this research, a systematic review has shown~\cite{Alene2021SerialMeta-analysis} that the incubation period pour COVID\nobreakdash-19 ranges from 2 to 14 days with a mean serial interval estimated to be 5.2. Hence, the decay of probability can only start from the earliest incubation possible and the choice of lambda should reflect that.

The public statistics from \cite{Foundation2021NHSTracker} gives us an indication on the potential ratio of potential positive tests versus negative tests outcomes linked to at least one close contact. Thus, we consider our data to be relatively balanced (we would select the same number tests which are coming out negative than positive), Accuracy is a proper metric to measure the performance of the models but additional such as F1 measure and AUC can give further insights on the performance of our model during the evaluation.

\subsection{Final model selection}\label{sec:model_selection}

\subsubsection{Performance metrics}
The evaluation of the trained models on the test data sets (from the 150k, 250k, 500, 750k and 1M records data sets) consistently ranked Ada Boost, Light Gradient Boosting Machine, Linear Discriminant Analysis, Logistic Regression and Ridge Regression as the best five performers with respect to the metrics chosen.
Table \ref{tab:performances} shows the average performances of the various models on the test data sets. The performances of the Light Gradient Boosting Machine, Logistic Regression and Linear Discriminant Analysis are overall extremely close, but we selected Light Gradient Boosting Machine for this project to be implemented in the enclave into the platform.

\subsubsection{Impact of the hyperparameters}
We looked at the impact of hyperparameters on the performances of the selected model. For this, we used 5-fold cross-validation and calculated optimised values for: a) the number of iterations, b) the maximum depth of each tree at each step, c) the learning rate, d) the maximum number of features. The tuning of the hyperparameters aimed to maximise the Accuracy score of the model on the training set. 

\subsubsection{Selection of the Lambda}
A small grid search of $\lambda$ between 0.00005 to 0.0005 with steps of size 0.0001, yielded $\lambda$ = 0.0001 as an adequate value for our proposed model.

\section{Discussion and Future Work}\label{sec:discussion}

The MOAI platform aims to provide a fully transparent implementation of practical contact tracing with research endpoints employing TEE as a base. Nevertheless, there are some identified gaps in the platform. We aim to address them as the next step in future work.

\topic{We explain the current limitation that because the platform is not open source that is still a limitation due to the implicit trust that the platform will do what it says}
The current MOAI implementation relies on an old version of the SCP that does not support the smart contract to be responsible for encrypting the data it exchanges with the mobile client application, the web portal or a third party auditing tool. As such, the Secretarium engine is implicitly trusted to handle decryption of the payload and offloading to the smart contract. This is a limitation of the current implementation as the code of the Secretarium engine is not publicly available.
Moreover, the code authenticity controls useful for checking the validity of the TEE's computation against the open-sourced smart contract is not immediately present and available from the Secretarium platform. Therefore, the general public cannot be sure the computation followed the prescription.
Work is on the way to enable a split header/payload encryption mechanism within the SCP protocol. Future work on MOAI to leverage this capability would address the limitation on trust and permit complete cryptographic control of the computation. Also, this work would enable proper authenticity validation checks.

It is worth mentioning that beyond the current technical limitations linked to the MOAI platform at the time of writing, there are intrinsic limitations to a unique ID venue-based tag scanning. It is possible to attack the system by copying QR Codes and placing copies in different locations or venues. While it would not render these type of systems ineffective, it can affect their performances the more such frauds are carried out. Being located in publicly available locations, there are no known mechanisms to prevent or fight such tempering.

\section{Conclusion}

\topic{We start by summarising the contributions and highlighting the challenges}
The approach presented in this paper is intended to propose a systematic approach to observational data collection at scale with the general public’s involvement without relying on sensors. Relying on laymen presents a challenge as the tools to enable the collection need to be readily available and the questionnaires easy to understand and quick to answer to encourage participants to contribute. Another key challenge is to demonstrate the trustworthiness of the proposed platform. The protection of everyone’s privacy while enabling researchers to carry out valuable research is paramount. One reason the first version of the NHS Test and Trace app failed to be adopted by the public was partially because of that lack of transparency on how the collected data would be processed~\cite{TheHealthFoundation2020GovernmentCOVID-19}, the security of the data (both in terms of who gets access to the data and under what conditions) and most importantly the absence of control over that data in the long run. MOAI aims to address those concerns by giving back control to the users and complete transparency on what data is collected, how it will be used, and whom.

\topic{We discuss the pertinence of the model and how it will evolve once we collect real data}
Finally, the proposed model relies on an extensive literature review available at the time. The synthetic data
provides insights on how some factors contribute to higher contamination but do not constitute definite proof. Nonetheless, it provides an indicator to the public and raises awareness on the risks of exposure in the context of this pandemic. The inclusion of that risk model in an easy to use environment could prove a valuable tool to encourage the adoption of safe behaviours. The continuous and safe collection of data will also enable the researchers to improve the proposed model in near real-time ultimately leading to better policies and a better understanding of the contamination patterns.

\section*{Acknowledgment}
The authors would like to thank the people at Secretarium Ltd for their support, especially Cecile Oblette, Lucy Pearson, Thomas Heinis, Pierre Richemond and Edward Boggis-Rolfe for their feedback, support and helpful comments. This work is supported by Innovate UK (Project Reference 72834) and Secretarium Ltd. 

\bibliographystyle{IEEEtran}
\bibliography{references.bib, other_references.bib}

\appendices
\section{Known risk factors}\label{app:Riskfactors}

Preliminary questions to be answered once in the app based on NHS's own classification.\\
\\
\textbf{Risk factors:}\\
Do you have any of the following risk factors:
\begin{itemize}
\item have had an organ transplant
\item are having chemotherapy or antibody treatment for cancer, including immunotherapy
\item are having an intense course of radiotherapy (radical radiotherapy) for lung cancer
\item are having targeted cancer treatments that can affect the immune system (such as protein kinase inhibitors or PARP inhibitors)
\item have blood or bone marrow cancer (such as leukaemia, lymphoma or myeloma)
\item have had a bone marrow or stem cell transplant in the past 6 months, or are still taking immunosuppressant medicine
\item have been told by a doctor you have a severe lung condition (such as cystic fibrosis, severe asthma or severe COPD)
\item have a condition that means they have a very high risk of getting infections (such as SCID or sickle cell)
\item are taking medicine that makes them much more likely to get infections (such as high doses of steroids or immunosuppressant medicine)
\item have a serious heart condition and are pregnant
\item are an adult with Down's syndrome
\item are an adult who is having dialysis or has severe (stage 5) long-term kidney disease
\item have been classed as clinically extremely vulnerable, based on clinical judgement and an assessment of your needs
\end{itemize}
a.	Yes\\
b.	No\\
If yes, then \textbf{High risk}
\\
\\
Do you have any of the following risk factors:
\begin{itemize}
\item are 70 or older
\item have a lung condition that's not severe (such as asthma, COPD, emphysema or bronchitis)
\item have heart disease (such as heart failure)
\item have diabetes
\item have chronic kidney disease
\item have liver disease (such as hepatitis)
\item have a condition affecting the brain or nerves (such as Parkinson's disease, motor neurone disease, multiple sclerosis or cerebral palsy)
\item have a condition that means they have a high risk of getting infections
\item are taking medicine that can affect the immune system (such as low doses of steroids)
\item are very obese (a BMI of 40 or above)
\item are pregnant
\item smoking
\end{itemize}
a.	Yes\\
b.	No\\
If yes, then \textbf{Moderate Risk}
\\
\\
Anything else can be classified as \textbf{Low Risk}

\section{NHS location types}\label{annex:NHSLocations}
NHS location types from https://www.gov.uk/create-coronavirus-qr-poster :
\begin{enumerate}
\item	Accommodation. For example, bed and breakfasts and campsites
\item	Childcare In public and private settings
\item	Education Including universities
\item	Events and conference space
\item	Finance and professional service. For example, high street banks and real estate agencies
\item	Medical facility. For example, hospitals, GP practices and veterinary clinics
\item	Non-residential institution. For example, community centres, libraries, crematoria and funeral homes
\item	Office location and workspace
\item	Personal care. For example, hair salons and barbers, spas and beauty salons
\item	Place of worship. For example, churches, synagogues, mosques, temples and meeting rooms
\item	Private event
\item	Recreation and leisure. For example, cinemas, theatres, museums and galleries
\item	Rental / hire locations
\item	Residential care. For example, care and nursing homes
\item	Restaurant, cafe, pub or bar
\item	Retail shops
\item	Sports and fitness facilities. For example, gyms, indoor sports facilities, swimming pools
\item	Transport For example, taxis and waiting rooms
\item	Other
\end{enumerate}

\section{MOAI Users Questionnaire}\label{app:UsersQuestionaire}
\textbf{Location types:}\\
Indoor: 1, 2, 3, 5, 6, 8, 9, 10, 11, 12, 14, 16, 18\\
Can be both inside or outside: 7, 13, 15, 17, 19\\
Question for location type that can be both inside or outside: \\
1)	What was the location type?\\
a.	Indoor\\
b.	Outdoor\\
\\
Questions for all locations\\
1)	How many people other than you do you estimate were present?\\
a.	0\\
b.	1-5\\
c.	5-10\\
d.	11+\\
\\
2)	How long did you stay at the location?\\
a.	5\\
b.	10\\
c.	15\\
d.	20\\
e.	30\\
f.	45\\
g.	1h\\
h.	2h\\
i.	2h+\\
\\
3)	Where people and staff wearing masks (do not display for venue 15)?\\
a.	Yes\\
b.	No\\
If no: was the staff wearing any form of PPE?\\
If yes: were people using the PPE correctly? (e.g. covering both the nose and mouth)\\
\\
4)	Were people following the social distancing rules?\\
a.	Yes\\
b.	No\\
\\
5)	Were additional protection were put in place (e.g. one-way systems, walled separators at tills, etc.)?\\
a.	Yes\\
b.	no\\
if Yes: can you please describe it in a few words\\
\\
6)	How many were in your party?\\
a.	Just me\\
b.	2\\
c.	2-4\\
d.	4+\\
\\
7)	Were all the member of your party from your household?\\
a.	Yes\\
b.	No\\
If No:\\
\\
8)	Were all the members from your support bubble?\\
a.	Yes\\
b.	No\\
\\
9)	How was the air flow?\\
a.	Well ventilated (doors or windows open, large inside space e.g. museums, etc.)\\
b.	No natural ventilation and air conditioning or heating was present and very likely to be working\\
c.	The air was circulating a lot from unknown source\\
d.	Confined space with no apparent ventilation\\
\\
10)	How was the temperature in the venue?\\
a.	Felt warm (above 25 degrees Celsius)\\
b.	Normal room temperature (between 20 and 25 degrees Celsius)\\
c.	Felt cold (below 19 degrees Celsius) \\
\\
11)	How was the humidity in the venue?\\
a.	Identical to outside \\
b.	Dryer than outside\\
c.	More humid than outside\\
\\
Question for venues 1, 4, 5, 8, 10, 15, 17, 18\\
1)	Were the surfaces cleaned after every usage?\\
a.	Yes\\
b.	No\\
c.	Often but not after every usage\\
\\
Questions for outdoor:\\
1)	Did any contact between members of the party occur during the gathering?\\
a.	Yes\\
b.	No\\
\\
2)	Did it involve singing or physical activities?\\
a.	Yes\\
b.	No\\

\end{document}